\documentclass[nolinenumbers,trackchanges]{aastex701}

\begin{document}

\title{Are Thalassa and Despina in Resonance Lock with Neptune's Oscillations?}

\author[orcid=0000-0003-1226-7960]{Matija \'Cuk}
\affiliation{SETI Institute}
\email[show]{mcuk@seti.org}  

\author[orcid=0000-0002-3544-298X]{Harrison F. Agrusa} 
\affiliation{Universit\'e C\^ote d'Azur, Observatoire de la C\^ote d'Azur, CNRS, Laboratoire Lagrange, Nice, France}
\affiliation{Centre national d'\'etudes spatiales (CNES), Paris, France}
\email{hagrusa@oca.eu}

\author[orcid=0009-0007-3493-2139]{Marina Brozovi\'c}
\affiliation{Jet Propulsion Laboratory}
\email{marina.brozovic@jpl.nasa.gov}

\author[orcid=0000-0002-8592-0812]{Matthew M. Hedman}
\affiliation{University of Idaho}
\email{mhedman@uidaho.edu}

\begin{abstract}
The two innermost moons of Neptune, Naiad and Thalassa, are currently in a 73:69 mean-motion resonance. This resonance relies on the large inclination of Naiad, and we estimate that Naiad requires multiple Gyr to reach its $4.7^{\circ}$ inclination through this resonance. However, we find through direct numerical simulations that the current Naiad-Thalassa resonance is unstable on Myr timescales due to perturbations from the neighboring moon Despina. As this instability is a product of convergent tidal evolution predicted by equilibrium tidal theory, we propose that the innermost moons of Neptune may migrate through resonant-lock tides. If both Despina and Thalassa are locked to two resonant oscillations modes within Neptune, the frequencies of which evolve approximately in parallel, Naiad-Thalassa resonance can be stable for much longer. We find that Lindblad resonances with low-order $l=m=1$, $n=1$ g-modes at Neptune may be suitable candidates for driving the resonant-lock evolution of Thalassa and Despina, and possibly even Galatea.
\end{abstract}

\section{Introduction}

Naiad, Thalassa and Despina are the innermost moons of Neptune and, together with Galatea, Larissa, Hippocamp and Proteus (see Table \ref{table}), they are thought to be the re-accreted remnants of Neptune's original satellite system that was destroyed during Triton's capture \citep{gol89, cuk05, agn06, ruf17}. The most remarkable orbital characteristic of the inner moons is the 4.7$^{\circ}$ inclination of Naiad, making it exceptional among the giant planets' close-in small moons. This inclination was previously thought to be a relic of a past interaction between Naiad and other moons \citep{ban92, zha08}, but \citet{bro20} found that Naiad and Thalassa are currently in a 73:69 mean-motion resonance (MMR). This MMR has the resonant argument $73 \lambda_{\theta}-69\lambda_{\nu}-4\Omega_{\nu}$, where $\lambda$ and $\Omega$ are the mean longitude and the longitude of ascending node, while subscripts $\nu$ and $\theta$ refer to Naiad and Thalassa, respectively. This is the only known stable fourth-order resonance among planetary satellites, as well as the second known inclination-type resonance \citep[the first being 4:2 MMR between Saturn's moons Mimas and Tethys;][]{md99}. 

Given that all of Neptune's inner moons other than Hippocamp and Proteus are interior to the synchronous orbit (i.e. they orbit faster than Neptune rotates), they are expected to migrate inward due to tidal interactions with Neptune. In classical tidal theory, the rate of recession due to tidal dissipation within the planet is \citep{md99}:
\begin{equation}
{\dot{a} \over a} = - {3 k_{2p} \over Q_p} {m_s \over m_p} \Bigl({R_p \over a}\Bigr)^5 n    
\label{equilibrium}
\end{equation}
where $a$ and $n$ are the moon's semimajor axis and mean motion, $m_s$ and $m_p$ the satellite's and the planet's masses, and $R_p$, $k_{2p}$ and $Q_p$ the planet's radius, tidal Love number and tidal quality factor, respectively. Therefore, tidal migration is expected to be proportional to the moon's mass, with the dependence on orbital distance being less relevant due to the inner moons' closely packed orbits (Table \ref{table}). In the classical paradigm, Thalassa is expected to migrate faster than Naiad, which would push Thalassa and Naiad deeper into their MMR and increase Naiad's inclination over time. 

Tidal evolution timescale depends on the ratio of Neptune's tidal quality factor $Q_N$ and tidal Love number $k_{2n}$. Using constraints from past resonances among inner moons (assuming equilibrium tides) \citet{zha08} find the range $2.2 \times 10^4 < Q_N/k_{2n} < 8.8 \times 10^4$ for this ratio. More recently, \citet{wan25} find, using the observed motion of Triton, that the minimum value of $Q_N/k_{2N}=8.2 \times 10^{4}$, but this result depends on using a priori constraints. For the \citet{wan25} value, Triton will fall into Neptune in about 28~Gyr, with the its lifetime being proportional to $Q_N/k_{2N}$, so frequency-independent values of $Q_N/k_{2N} < 10^4$ would require a relatively short future lifetime of Triton.


\begin{deluxetable*}{lcrrr}
\tablewidth{0pt}
\tablecaption{Physical and orbital parameters of the Neptune's moons out to Triton. Orbits: from \cite{jac09} for Triton, \cite{bro20} for others. Radii: from \cite{dav91} for Triton, \cite{sho19} for Hippocamp, and \cite{kar03} for others. Free inclination is measured relative to the local Laplace plane, which is itself determined by the sum of outside perturbations.\label{table}}
\tablehead{
\colhead{Moon} & \colhead{Radius [km]} & \colhead{Semimajor axis [km]} & \colhead{Eccentricity} & \colhead{Free Inclination [$^{\circ}$]}
}
\startdata
Naiad & $33\pm3$ &  48228. & 0.00014 & 4.728\\
Thalassa & $41\pm3$ & 50075. & 0.00019 & 0.168\\
Despina & $75\pm3$ &  52526. & 0.00027 & 0.039\\
Galatea & $88\pm4$ & 61953. & 0.00020 & 0.010\\
Larissa & $97\pm3$ &  73548. & 0.00121 & 0.214\\
Hippocamp & $17\pm2$ & 105283. & 0.00001 & 0.002\\
Proteus & $210\pm7$ &  117647. & 0.00047 & 0.042\\
Triton & $1352.6\pm2.4$ & 354759. & 0.00001 & 156.865\\
\enddata
\end{deluxetable*}

Apart from its orbital inclination, Naiad is remarkable for its seemingly precarious physical integrity. Naiad may be interior to its Roche limit, depending on its still poorly-known density \citep{tis13, agr25}. This may potentially indicate that Naiad is at least partially held together by cohesive forces,  rather than gravity alone. Furthermore, Naiad's high inclination puts it at risk from ``sesquinary catastrophe'' \citep{cuk23, agr25}, in which the orbits of ejecta from a moon precess out of alignment with the moon's own orbit, leading to potentially erosive re-impacts. In order to avoid sesquinary catastrophe, Naiad is almost certainly not be a rubble pile \citep{agr25}. This would contrast Naiad (and presumably its neighbors) to the inner moons of Uranus, which are likely to be rubble piles \citep{tis13, hed15}\footnote{A rubble pile is defined by being a gravitational aggregate and not necessarily a low-density body. Uranian moon Cordelia needs a density above $1000$~kg~m$^{-3}$ to be gravitationally bound, in agreement with observations \citet{fre24}, and consistent with it being an aggregate of rock-rich material.}. The inner moons of Uranus are known to be dynamically unstable and short-lived \citep{dun97, fre12, cuk22b}, motivating the question of the age of Naiad and other inner moons of Neptune. If the Naiad-Thalassa resonance existed for billions of years, that would also indicate that Naiad and Thalassa themselves are similarly old.

In this Letter we will first offer some basic analysis of the Naiad-Thalassa resonance in isolation, mostly confirming theoretical expectations \citep[e.g.][]{md99} and past work \citep{bro20}. We will then show that the Naiad-Thalassa resonance is strongly affected by the presence of Despina, with important implications for either the longevity of the resonance or the nature of the moons' tidal evolution. Finally, in Section \ref{sec:modes} we will propose a tentative hypothesis that Thalassa and Despina maybe in resonance lock \citep{ful16} with the internal oscillations of Neptune. 

\section{Naiad-Thalassa Resonance in Isolation}

\begin{figure}
\plotone{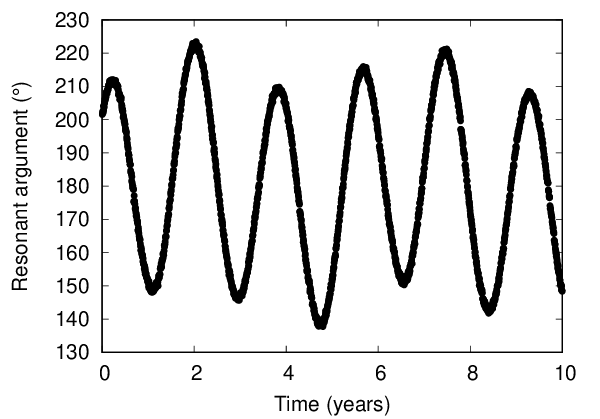}
\caption{The resonant argument of the Naiad-Thalassa resonance using the initial vectors from the JPL's Horizons ephemeris system, which are based on the solution presented by \citep{bro20}. \label{resarg}}
\end{figure}

The first step in modeling the Naiad-Thalassa resonance is setting up initial conditions for the moons, as well as orientation of Neptune's spin axis. We downloaded initial conditions for Naiad, Thalassa, Despina and Triton fom the JPL HORIZONS site\footnote{The web address is https://ssd.jpl.nasa.gov/horizons/ . We accessed the site on April 16, 2025 and used January 1, 2020 as our epoch.}, but Despina was omitted from  simulations shown in this section. The masses for Naiad and Thalassa were assumed to be $1.2 \times 10^{17}$~kg and $3.56 \times 10^{18}$~kg, based on \citep{bro20}. We use the numerical integrator {\sc simpl} \citep{cuk16}, which is based on the symplectic algorithm of \citet{cha02}. Figure \ref{resarg} shows the resonant argument of the Naiad Thalassa resonance $73\lambda_{\theta}-69 \lambda_{\nu} - 4 \omega_{\nu}$ over 10 years in the resulting simulation. Our model is therefore consistent with the solution of \citet{bro20} but not independent, as the vectors we used are derived from their work. The resonant argument shows approximately sinusoidal librations with a semi-amplitude of up to 50$^{\circ}$, indicative of a stable mean-motion resonance.

As we mention in the Introduction, we expect the orbits of Thalassa and Naiad to converge due to tidal dissipation inside of Neptune. As the two moons are in a resonance, their semimajor axes should evolve in lockstep, with the inclination of Naiad growing over time. For determining the system's dynamical history, it would be useful to estimate the amount of orbital evolution necessary to produce the current inclination of Naiad. \citet{md99} show how this can be done for Titan-Hyperion eccentricity-type resonance, and we use the same approach for the Naiad-Thalassa resonance by replacing $e$ by $\sin{i}$. Equation 8.423 in \citet{md99} modified in this manner is:
\begin{equation}
{d(\sin{i}) \over \sin{i} dt} = {1 \over \sin^2{i}}{j_5 \over 2 j_2} {\dot{a} \over a}
\end{equation}
where $j_2=73$ and $j_5=-4$ for the Naiad-Thalassa 73:69 MMR. When we integrate this equation, assuming a much smaller initial inclination, we get
\begin{equation}
\sin^2{i} = - {4 \over 73} {\Delta{a} \over a}
\label{deltaa}
\end{equation}
Substituting the current inclination of Naiad from Table \ref{table}, we get that $\Delta{a}/a=-0.12$. This calculation assumed that Thalassa is the only moon experiencing tidal evolution, but in reality Naiad itself would be migrating about half as fast. As the inclination increase depends on how fast two orbits converge, rather than on their absolute tidal evolution, we must multiply the above estimate by a factor of two. Therefore, simple analytical estimates suggest that the orbits of Thalassa and Naiad have shrunk by about 20\% over the whole lifetime of the resonance, assuming that most of Naiad's inclination was acquired in the present resonance. 

\begin{figure}
\plotone{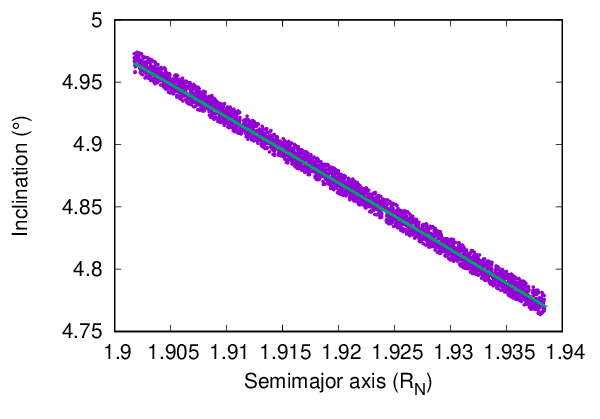}
\caption{Purple: The increase of Naiad's inclination as a function of its semimajor axis in a $3 \times 10^5$~yr simulation that included only Naiad and Thalassa, started from the present day, and used $Q/k_2=33$ for Neptune (i.e. it was accelerated by at least a factor of 1000). Green: A simple approximation of this dependence as $i = 7.195^{\circ} \sqrt{2.378-a}   $\label{a-i}}
\end{figure}

We can also estimate the amount of past tidal evolution directly from numerical simulations. Figure \ref{a-i} plots the results of a greatly accelerated simulation of future orbital evolution of Naiad and Thalassa (without other moons, with an exception of Triton). Since theory suggests that the square of sine of inclination is proportional to the semimajor axis change, we fit the resulting curve with a function $i=i_o * \sqrt{a_o-a}$, where $a_0$ is the semimajor axis at resonance capture and $i_0$ is a constant (we ignore the difference between $i$ and $\sin{i}$ here). The best fit suggests $a_0=2.38 R_N$, implying $\Delta a/a=0.23$, in line with our analytical estimates.

For $Q_N/k_{2N} =3.3 \times 10^{4}$, tidal migration amounting to 20\% of Thalassa's semimajor axis would take longer than the age of the Solar System. Additionally, Naiad and Thalassa should fall into Neptune about 2 Gyr from now, due to the steep dependence of tidal evolution on distance. A somewhat more dissipative Neptune in the past could provide enough tidal migration to generate Naiad's inclination over the age of the Solar System, but we find it more likely that Naiad may have been in other resonances in the past and/or tidal evolution is not governed by Eq. \ref{equilibrium}. We first address the first possibility, while the second is addressed in the next section.

While the Naiad-Thalassa resonance is currently stable (in isolation) and leads to an increase in Naiad's inclination, there is an open question of how this resonance originated. In general, two-body MMR capture is possible only if the relevant orbital element (here the inclination of Naiad) is small at the time of capture, and the convergent evolution of the two bodies is slow. We attempted to simulate capture by using extremely slow orbital migration of Thalassa (with $Q_N/k_{2N} > 10^5$) and low  initial inclination $i_{\nu} \simeq 10^{-4 \circ}$, with no success. Furthermore, no ``kick'' or other indication of the resonance presence was detected, suggesting that the resonance is extremely weak for small inclinations of Naiad. This is unsurprising, given that this resonance's strength is proportional to $i_{\nu}^4$. It was previously proposed by \cite{bro20} that Naiad must have been captured into the current resonance after its inclination was already excited, and our preliminary simulations support that idea.  

While the inclination of Naiad could not have been increased from zero to the present value solely through the current 73:69 resonance with Thalassa (as resonant capture at $i=0$ is not possible), the estimated amount of orbital migration required for inclination growth is approximately applicable to other resonances. Eq. \ref{deltaa} would give us similar order of magnitude for $\Delta{a}$ required for producing the same inclination, as long as the perturbing moon has mean motion that is not too different from that of Naiad (which would apply to Despina, for example). While a Despina-Naiad resonance would increase Naiad's inclination faster than the current Naiad-Thalassa MMR, we expect the perturbations from Thalassa to limit the long-term  stability of such a resonance (see next Section).

\section{Influence of Despina}

\begin{figure}[h!]
\plotone{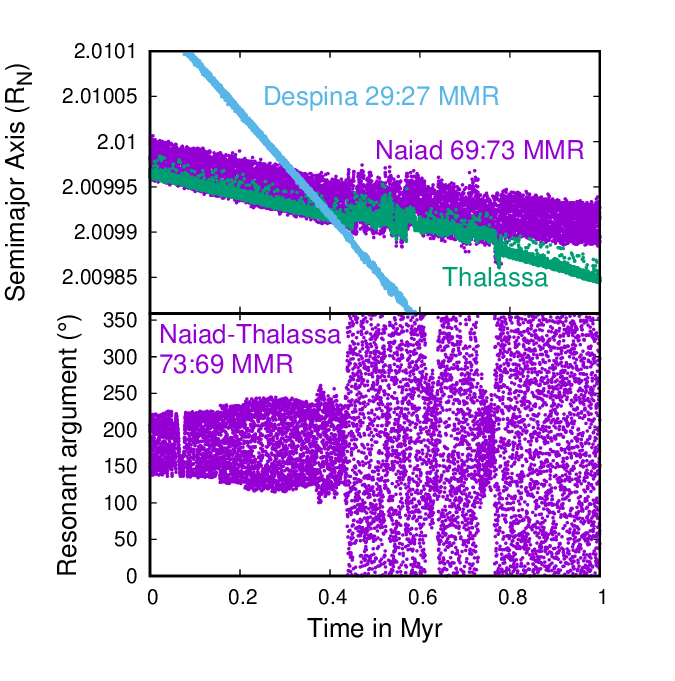}
\caption{\sf A simulation of future evolution of orbits of Naiad, Thalassa and Despina assuming equilibrium tides within Neptune with a constant $Q_N/k_{2N}=3.3 \times 10^4$. The top panel shows the semimajor axis of Thalassa (green) as well as the location of the currently active 69:73 MMR with Naiad (purple; approximately $(73/69)^{2/3} a_{\nu}$) and the 29:27 resonance with Despina (cyan; approximately $(27/29)^{2/3} a_{D}$, the resonant argument is $29 \lambda_D - 27\lambda_{\theta} - 2\Omega_{\theta}$). The bottom panel show the resonant argument $73 \lambda_{\theta}-69\lambda_{\nu}-4\Omega_{\nu}$ of the Naiad-Thalassa 73:69 resonance. Naiad-Thalassa resonance is perturbed about 430 kyr into the simulation due to Despina's resonant perturbations, and finally breaks around 800 kyr.}\label{qtides}
\end{figure}

In Fig. \ref{a-i} we plot long term evolution of the Naiad-Thalassa resonance in isolation, i.e. without the influence of any of the other inner moons of Neptune (only other perturbers were Triton and the Sun). However, as Despina orbits very close to Thalassa, it is expected that Despina may have significant dynamical influence on the two smaller moons. Figure \ref{qtides} plots a simulation of the Myr-scale future evolution of the three moons, using the integrator {\sc simpl} and assuming equilibrium tides with $Q_{N}/k_{2N}=3.3 \times 10^4$ for Neptune.

Figure \ref{qtides} demonstrates that the current Naiad-Thalassa resonance breaks in less than 1~Myr due to its interaction with Despina. As in previous runs, this integration includes the full effects of Triton and the Sun, which only act to slightly alter the Laplace planes of the inner satellites. Our choice for Neptune's tidal properties ($Q_N/k_{2N}=3.3 \times 10^4$) is on the low (i.e. faster evolution) end of the plausible range of $2.2-8.8 \times 10^4$ found by \cite{zha08}, but even the slowest expected orbital evolution would result in Naiad-Thalassa resonance breaking in just over 1~Myr. It appears that the Naiad-Thalassa resonance is broken due to Thalassa and Despina encountering their common 29:27 MMR (light blue line in Fig. \ref{qtides}). There is no reason to think that this particular resonance with Despina is unique, and this just illustrates that convergent evolution of the Thalasa-Naiad pair with Despina will encounter a disruptive resonance relatively quickly. If correct, such a short lifetime of the current Naiad-Thalassa 73:69 MMR would make it extremely unlikely for humans to observe this resonance. 


\begin{figure}[h!]
\plottwo{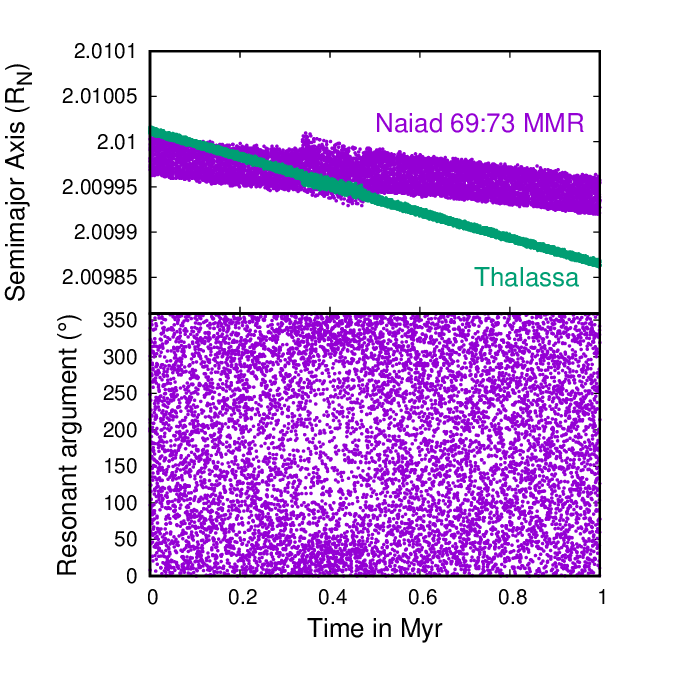}{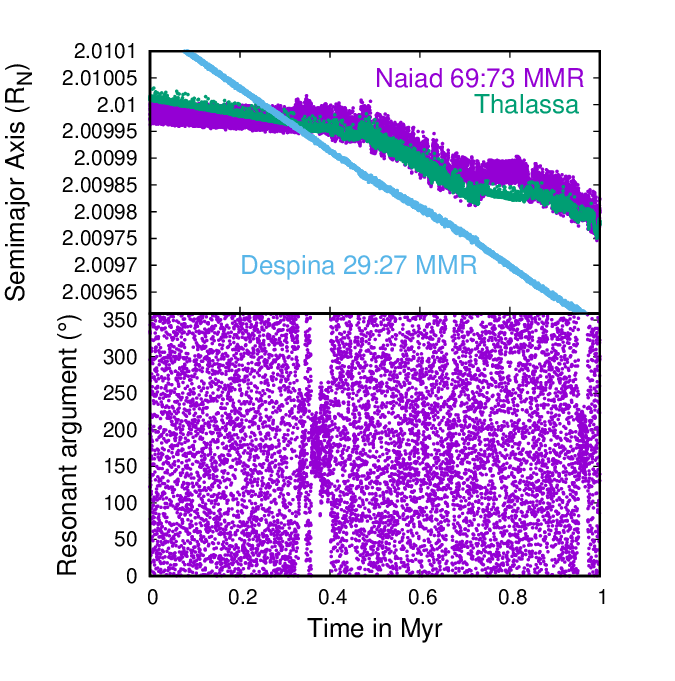}
\caption{\sf Similar to Fig. \ref{qtides}, but with Thalassa shifted to a slightly larger semimajor axis, outside of the resonance with Naiad. The left hand simulations did not include Despina, while the right-hand one did. The top panels show the semimajor axis of Thalassa (green) as well as the location of the currently active 69:73 MMR with Naiad (purple) and (in the right-hand plot) the 29:27 resonance with Despina (cyan; the resonant argument is $29 \lambda_D - 27\lambda_{\theta} - 2\Omega_{\theta}$). The bottom panels show the resonant argument $73 \lambda_{\theta}-69\lambda_{\nu}-4\Omega_{\nu}$ of the Naiad-Thalassa 73:69 resonance. The two-body simuations shows a simple resonance crossing during which the resonant argument avoids the librationcenter. In the three-body simulation there is also no permanent capture into any resonance, but chaotic interactions between three moons lead to temporary periods of libration in resonance for Naiad and Thalassa.}\label{crossing}
\end{figure}

Before considering solutions to the apparent short lifetime of the Naiad-Thalassa resonance, we will consider how Despina may have affected the origin of the current Naiad-Thalassa MMR. Considered in isolation, Naiad and Thalassa cannot enter their present MMR at the present inclination of Naiad, as their convergent migration would lead to a jump through the resonance (Fig. \ref{crossing}, left-hand side). However, the presence of Despina's numerous MMRs with both Naiad and Thalassa makes the Naiad-Thalassa interaction chaotic (Fig. \ref{crossing}, right-hand panels). We find that at least temporary resonance capture is possible even if the initial inclination of Naiad is high, but that the current Myr-term stable Naiad-Thalassa MMR does not appear to be a common outcome. Clearly, large-scale numerical simulations are needed to properly explore this stochastic process. It is in principle plausible that during this long-term chaotic evolution of the three moons there may be periods of relatively stable two-body resonances. The high inclination of Naiad makes even the high-order resonances such as the current 73:69 MMR at least temporarily viable.

While it is conceivable that the observed Naiad-Thalassa resonance is just a temporary stage of a long-term chaotic evolution, there are also simple analytical arguments against this hypothesis. The inclination of Naiad is about thirty times larger than that of Thalassa, although Thalassa is likely only around three times more massive. A fully chaotic evolution is expected to produce an approximate equipartition of eccentricities and inclinations among the participating moons. Eccentricities are likely to be quickly damped through satellite tides for such close-in moons \citep{md99}, while inclinations should be long-lived for all inner moons \citep[cf. ][]{che14}. Therefore, while the long-term chaotic interactions must be modeled directly, there are good reasons to think that the observed resonance between Naiad and Thalassa may actually be a long-term stable state, and we address one such possibility in the next section.

Before moving on to more exotic explanations, it is important to ask if there are any unrecognized resonances in the system that could explain the stability of the Naiad-Thalassa resonance. While it is certain that there are no two-body resonances between Thalassa and Despina, it is conceivable that these two moons could be part of  a three-body resonance, which could make them evolve in unison. We have explored all conceivable first and second order three-body resonances (the former being of the eccentricity type, the latter also involving inclinations) using mean motions and precessions from \citet{bro20}, and we found none (M. Brozovi\'c and M.\'Cuk conducted independent searches). Furthermore, any unseen long-term stable resonance would have to increase eccentricity or inclination of one of the moons over time, and none of the inclinations or eccentricities appear unusually high, especially in comparison with Naiad's inclination. While it is hard to prove a negative, and there could exist never-before seen types of MMRs, we must conclude for now that additional currently active resonances between the moons are unlikely to solve the problem of the stability of the Naiad-Thalassa resonance. However, past three-body resonances are certainly a possibility and could have contributed to the current inclination of Naiad. 

\section{Discussion: Resonance-Lock Tides at Neptune?} \label{sec:modes}

The observed fast tidal migration of Rhea and Titan \cite{lai17, lai20} is best explained by the mechanism of resonance-lock tides in giant planets. Unlike equilibrium tides, in which the planet's response is only a weak function of a moon's orbital frequency, resonance-lock tides assume that there are certain resonant frequencies at which the planet dissipates much more energy than elsewhere. These resonances are relatively narrow but shift in frequency space as the planet's interior structure evolves. If the background (i.e. out of resonance) tidal evolution rate of a satellite's orbital frequency is slower than the shifting of the resonant frequency itself, the satellite will ``lock'' to the resonant frequency, and its orbital frequency will track the tidal resonant frequency \cite{ful16}. Resonance-lock requires that the resonant locations and satellite orbits migrate in the same direction, which is outward for moons outside the synchronous radius (such as Saturn's Rhea and Titan) and inward for moons inside the synchronous orbit (such as Thalassa and Despina).  

If a planet has a number of tidal resonances with internal oscillations, then multiple moons can be resonance-locked at the same time. In this case, the tidal evolution of each moon would depend solely on the shifting of the planet's own frequencies, and not on the moons' own parameters such as their masses. If two planetary tidal resonances are evolving in parallel, resonance-lock enables two moons of dissimilar masses to evolve on parallel (or nearly parallel) trajectories, making MMR encounters rare, or avoiding them altogether. On the other hand, if a resonance-locked moon encounters a mean-motion resonance with a non-resonance locked moon, mean-motion resonance can be established as in the classical case (usually requiring convergent orbits and a low initial eccentricity or inclination). The pair of moons in the MMR will subsequently migrate at the rate of the shifting planetary frequency that the trailing moon is resonance-locked to. 

There is a clear potential relevance of the resonance-lock mechanism for explaining the apparent stability of the Naiad-Thalassa resonance. If both Despina and Thalassa are resonance-locked to two different internal oscillation modes of Neptune that are evolving at similar rates, the orbits of Despina, Thalassa and Naiad (locked to Naiad through the conventional MMR) could evolve in parallel, without crossing any new mutual MMRs. These proposed tidal resonances would have to move inward in order for resonance-lock tides to work on the inner moons of Neptune. 


\begin{deluxetable*}{lrrr}
\tablewidth{0pt}
\tablecaption{Locations of Lindblad and corotation resonances of $l=m \le 10$, $n=1$ g-modes of Neptune as calculated by \citet{ahe22}. \label{ahearn}}
\tablehead{
\colhead{m} & \colhead{Pattern freq. [$^{\circ}$ day$^{-1}$]} & \colhead{Lindblad res. [km]} & \colhead{Corot. res. [km]}}
\startdata
1&1808.1-34.7&58,959 + 766&37,170 + 481\\	
2&1683.1-28.2&51,058 + 578&38,986 + 441\\
3&1566.6-24.6&49,518 + 526&40,893 + 433\\	
4&1462.1-21.9&49,668 + 502&42,815 + 432\\	
5&1369.0-19.7&50,502 + 489&44,732 + 432\\	
6&1287.1-17.7&51,646 + 480&46,611 + 432\\	
7&1215.6-16.1&52,917 + 472&48,417 + 431\\	
8&1153.8-14.7&54,217 + 465&50,129 + 429\\	
9&1100.5-13.5&55,495 + 458&51,736 + 426\\	
10&1054.3-12.5&56,722 + 451&53,235 + 423\\	
\enddata
\end{deluxetable*}

While the identity of planetary oscillation modes that drive the apparent resonance-lock of Rhea and Titan is unknown, there are a number of previously predicted normal modes of Saturn that have been detected through their perturbations on Saturn's rings \citet{hed13, hed14, fre19}. These are mostly f-modes \citep{mar93, ful14, dew21, man21}, and at Saturn they are perturbing the ring particles' orbits at specific planetocentric distances. It is intriguing to consider whether any of these known types of oscillations could drive resonance-lock tidal evolution if they were to interact with relatively massive moons rather than low-mass ring particles.

\begin{figure}
\plotone{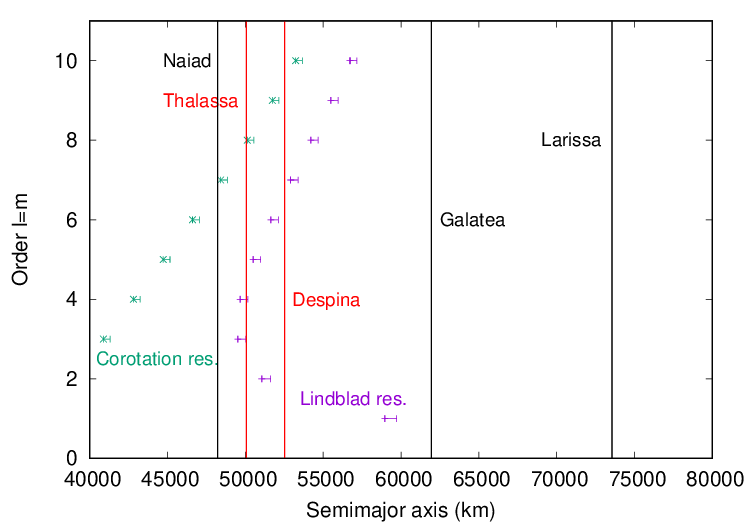}
\caption{Comparison of locations of $n=1$, $l=m$ g-mode resonances (up to order $l=m=10$) calculated by \citep[Table \ref{ahearn}; ][]{ahe22} and the semimajor axes of the inner moons of Neptune (Table \ref{table}). Locations of Lindblad resonance are plotted in purple and those of corotation resonances in green. Semimajor axes of moons suspected to be in resonance lock are plotted in red.  \label{modes_plot}}
\end{figure}

\citet{ahe22} have modeled locations of resonances f- and g-modes around Uranus and Neptune, looking for potentially observable effects on the rings. What is of interest is that their predicted locations of locations of Lindblad resonances with Neptune's low-order $n=1, l=m$ modes are at, or close, to the orbits of Thalassa and Despina \citep[Table \ref{ahearn}, showing results from][]{ahe22}, offering a candidate planetary oscillation for the resonance-locks with these moons (also Fig. \ref{modes_plot}). We find the tentative correspondence between the locations of the g-mode resonances and the orbits of the apparently resonance-locked Thalassa and Despina to be extremely interesting. Thalassa's semimajor axis is just outside the predicted error bars for $l=3$ mode and inside the error bars for the $l=4$ mode. Despina is about one thousand kilometers outside the error bars for the $l=m=2$, $n=1$ g-mode. 

We have not yet conducted any modeling of the interaction of a massive moon with a g-mode, and we are not aware of such work in the literature. Our argument for the viability of this mechanism is one of symmetry: if internal oscillation resonances can perturb the orbiting particles, then satellites should be able to perturb the planetary oscillation modes. Resonant response by the planet may lead to greatly enhanced amplitude of the oscillation that would also produce enhanced dissipation. As the energy for this interaction is ultimately supplied by the satellites' orbits, orbital evolution has to result from this process. Additionally, as these are Lindblad-type resonances, the satellite's eccentricities also may be damped by the planets resonant response faster than they would be damped by dissipation within the moons. The question of how the present configuration of Despina, Thalassa and Naiad is established in the paradigm of resonance-lock tides will be the focus of our future work. 

It is worth noting some important differences between these potential planet-satellite interactions at Neptune and those proposed to explain the rapid orbital evolution of Saturn's satellites. Most obviously, Thalassa and Despina are orders of magnitude less massive than any of Saturn's mid-sized moons. It is not yet clear whether such small moons can excite planetary normal modes enough to achieve the resonance-lock state. However, the above analysis also indicates that Thalassa and Despina could be in resonance with low-order g modes. These particular modes are expected to have very low dissipation rates \citep{ful26}, and this is observed to be the case for Saturn \citep{hed13}. If the quality factor of these modes is high enough, that could allow them to be excited by smaller moons sufficiently efficiently to achieve the resonance-lock state. Future work may also need to consider the relative importance of the satellites' gravitational perturbations and the planet's internal processes for exciting and aligning these modes, as well as how the frequencies of these particular modes would change relative to each other as the planet evolves.

The predicted locations of the $m=l$, $n=1$ g-mode Lindblad resonances at Neptune open another possibility that goes beyond what has been suggested by our numerical simulations. Despina is about one thousand kilometers outside the predicted range of locations for the $l=2$ g-mode Lindblad resonance. If we accept that there may be a mismatch of this magnitude between the predicted and actual locations of g-mode resonances, then it is possible that the next moon Galatea is in $l=m=1$, $n=1$ g-mode Lindblad resonance, as its semimajor axis is about 2,300~km outside of the predicted resonance range. This way Galatea, Despina and Thalassa would be resonance-locked to the three strongest Lindblad resonances with $l=m$, $n=1$ g-modes. We recognize that this hypothesis is highly speculative at this point, but we argue that is certainly deserving of further investigation.

\section{Conclusions} 

Using direct numerical simulations we find that the observed Naiad-Thalassa 73:69 mean-motion resonance has a lifetime on the order of 1~Myr due to the perturbations of the neighboring moon Despina. This lifetime is in direct conflict with the time needed for the resonance to excite Naiad's inclination to its present value, which is estimated to be several Gyr. This conflict between inclination-growth timescale within the Naiad-Thalassa resonance and the lifetime of the resonance against Despina's perturbations is not dependent on the value of tidal $Q_N$, but is intrinsic to the model of equilibrium tidal evolution. We conclude that the most likely solution to this paradox is that the tidal evolution of some of the innermost moons of Neptune may not be governed by equilibrium tides, for which a satellite's tidal migration rate is directly proportional to its mass.

We propose that the anomalous properties of the Naiad-Thalassa resonance can be explained by resonance-lock tides within Neptune acting on Thalassa and Despina. This parallel evolution could explain the longevity of the Naiad-Thalassa resonance, as Despina would essentially evolve in parallel with those two moons instead of crossing MMRs with them. Particularly exciting is the possibility that the internal modes involved in Thalassa and Despina's resonance-lock are g-modes of Neptune \citep{ahe22}, enabling us to constrain the interior properties of the planet if this idea is correct. If the tidal evolution of Despina and Thalassa is driven by g-modes, we suspect on the basis of g-mode resonance locations that Neptune's moon Galatea may also be resonance-locked to a g-mode of Neptune.

\begin{acknowledgments}
M\' C is supported by NASA Solar System Workings award 80NSSC24K1842. HA acknowledges financial support from the Centre national d’études spatiales (CNES), France (ROR: https://ror.org/04h1h0y33) within the framework of the MMX mission.
M.B. research is conducted at the Jet Propulsion Laboratory, California Institute of Technology, under contract with the National Aeronautics and Space Administration (80NM0018D0004)
\end{acknowledgments}

\begin{contribution}

M\'C proposed the initial hypothesis and conducted the numerical simulations. All of the authors contributed to the scientific content and writing of the paper.


\end{contribution}

\bibliography{naiad_refs}{}

\begin{thebibliography}{}
\expandafter\ifx\csname natexlab\endcsname\relax\def\natexlab#1{#1}\fi
\providecommand{\url}[1]{\href{#1}{#1}}
\providecommand{\dodoi}[1]{doi:~\href{http://doi.org/#1}{\nolinkurl{#1}}}
\providecommand{\doeprint}[1]{\href{http://ascl.net/#1}{\nolinkurl{http://ascl.net/#1}}}
\providecommand{\doarXiv}[1]{\href{https://arxiv.org/abs/#1}{\nolinkurl{https://arxiv.org/abs/#1}}}

\bibitem[{C.~B. {Agnor} \& D.~P. {Hamilton}(2006){Agnor} \& {Hamilton}}]{agn06}
{Agnor}, C.~B., \& {Hamilton}, D.~P. 2006, \bibinfo{title}{{Neptune's capture
  of its moon Triton in a binary-planet gravitational encounter},} Nature, 441,
  192, \dodoi{10.1038/nature04792}

\bibitem[{H. {Agrusa} {et~al.}(2025){Agrusa}, {{\'C}uk}, {Nesvorny}, \&
  {Marschall}}]{agr25}
{Agrusa}, H., {{\'C}uk}, M., {Nesvorny}, D., \& {Marschall}, R. 2025,
  \bibinfo{title}{{The Curious Case of Neptune's Naiad},} in EPSC-DPS Joint
  Meeting 2025, Vol. 2025, EPSC--DPS2025--816, \dodoi{10.5194/epsc-dps2025-816}

\bibitem[{J.~A. {A'Hearn} {et~al.}(2022){A'Hearn}, {Hedman}, {Mankovich},
  {Aramona}, \& {Marley}}]{ahe22}
{A'Hearn}, J.~A., {Hedman}, M.~M., {Mankovich}, C.~R., {Aramona}, H., \&
  {Marley}, M.~S. 2022, \bibinfo{title}{{Ring Seismology of the Ice Giants
  Uranus and Neptune},} PSJ, 3, 194, \dodoi{10.3847/PSJ/ac82bb}

\bibitem[{D. {Banfield} \& N. {Murray}(1992){Banfield} \& {Murray}}]{ban92}
{Banfield}, D., \& {Murray}, N. 1992, \bibinfo{title}{{A dynamical history of
  the inner Neptunian satellites},} Icarus, 99, 390,
  \dodoi{10.1016/0019-1035(92)90155-Z}

\bibitem[{M. {Brozovi{\'c}} {et~al.}(2020){Brozovi{\'c}}, {Showalter},
  {Jacobson}, {French}, {Lissauer}, \& {de Pater}}]{bro20}
{Brozovi{\'c}}, M., {Showalter}, M.~R., {Jacobson}, R.~A., {et~al.} 2020,
  \bibinfo{title}{{Orbits and resonances of the regular moons of Neptune},}
  Icarus, 338, 113462, \dodoi{10.1016/j.icarus.2019.113462}

\bibitem[{J.~E. {Chambers} {et~al.}(2002){Chambers}, {Quintana}, {Duncan}, \&
  {Lissauer}}]{cha02}
{Chambers}, J.~E., {Quintana}, E.~V., {Duncan}, M.~J., \& {Lissauer}, J.~J.
  2002, \bibinfo{title}{{Symplectic Integrator Algorithms for Modeling
  Planetary Accretion in Binary Star Systems},} Astronomical Journal, 123, 2884

\bibitem[{E.~M.~A. {Chen} {et~al.}(2014){Chen}, {Nimmo}, \&
  {Glatzmaier}}]{che14}
{Chen}, E.~M.~A., {Nimmo}, F., \& {Glatzmaier}, G.~A. 2014,
  \bibinfo{title}{{Tidal heating in icy satellite oceans},} Icarus, 229, 11,
  \dodoi{10.1016/j.icarus.2013.10.024}

\bibitem[{M. {{\'C}uk} {et~al.}(2016){{\'C}uk}, {Dones}, \&
  {Nesvorn{\'y}}}]{cuk16}
{{\'C}uk}, M., {Dones}, L., \& {Nesvorn{\'y}}, D. 2016,
  \bibinfo{title}{{Dynamical Evidence for a Late Formation of
  Saturn{\textquoteright}s Moons},} Astrophysical Journal, 820, 97,
  \dodoi{10.3847/0004-637X/820/2/97}

\bibitem[{M. {{\'C}uk} {et~al.}(2022){{\'C}uk}, {French}, {Showalter},
  {Tiscareno}, \& {El Moutamid}}]{cuk22b}
{{\'C}uk}, M., {French}, R.~S., {Showalter}, M.~R., {Tiscareno}, M.~S., \& {El
  Moutamid}, M. 2022, \bibinfo{title}{{Cupid is not Doomed Yet: On the
  Stability of the Inner Moons of Uranus},} Astronomical Journal, 164, 38,
  \dodoi{10.3847/1538-3881/ac745d}

\bibitem[{M. {{\'C}uk} \& B.~J. {Gladman}(2005){{\'C}uk} \& {Gladman}}]{cuk05}
{{\'C}uk}, M., \& {Gladman}, B.~J. 2005, \bibinfo{title}{{Constraints on the
  Orbital Evolution of Triton},} ApJL, 626, L113, \dodoi{10.1086/431743}

\bibitem[{M. {{\'C}uk} {et~al.}(2023){{\'C}uk}, {Hamilton}, {Minton}, \&
  {Stewart}}]{cuk23}
{{\'C}uk}, M., {Hamilton}, D.~P., {Minton}, D.~A., \& {Stewart}, S.~T. 2023,
  \bibinfo{title}{{Sesquinary Catastrophe for Close-in Moons with Dynamically
  Excited Orbits},} ApJ, 957, 62, \dodoi{10.3847/1538-4357/acf613}

\bibitem[{M.~E. {Davies} {et~al.}(1991){Davies}, {Rogers}, \& {Colvin}}]{dav91}
{Davies}, M.~E., {Rogers}, P.~G., \& {Colvin}, T.~R. 1991, \bibinfo{title}{{A
  control network of Triton},} JGR, 96, 15675, \dodoi{10.1029/91JE00976}

\bibitem[{J.~W. {Dewberry} {et~al.}(2021){Dewberry}, {Mankovich}, {Fuller},
  {Lai}, \& {Xu}}]{dew21}
{Dewberry}, J.~W., {Mankovich}, C.~R., {Fuller}, J., {Lai}, D., \& {Xu}, W.
  2021, \bibinfo{title}{{Constraining Saturn's Interior with Ring Seismology:
  Effects of Differential Rotation and Stable Stratification},} \psj, 2, 198,
  \dodoi{10.3847/PSJ/ac0e2a}

\bibitem[{M.~J. {Duncan} \& J.~J. {Lissauer}(1997){Duncan} \&
  {Lissauer}}]{dun97}
{Duncan}, M.~J., \& {Lissauer}, J.~J. 1997, \bibinfo{title}{{Orbital Stability
  of the Uranian Satellite System},} Icarus, 125, 1,
  \dodoi{10.1006/icar.1996.5568}

\bibitem[{R.~G. {French} {et~al.}(2024){French}, {Hedman}, {Nicholson},
  {Longaretti}, \& {McGhee-French}}]{fre24}
{French}, R.~G., {Hedman}, M.~M., {Nicholson}, P.~D., {Longaretti}, P.-Y., \&
  {McGhee-French}, C.~A. 2024, \bibinfo{title}{{The Uranus system from
  occultation observations (1977-2006): Rings, pole direction, gravity field,
  and masses of Cressida, Cordelia, and Ophelia},} \icarus, 411, 115957,
  \dodoi{10.1016/j.icarus.2024.115957}

\bibitem[{R.~G. {French} {et~al.}(2019){French}, {McGhee-French}, {Nicholson},
  \& {Hedman}}]{fre19}
{French}, R.~G., {McGhee-French}, C.~A., {Nicholson}, P.~D., \& {Hedman}, M.~M.
  2019, \bibinfo{title}{{Kronoseismology III: Waves in Saturn's inner C ring},}
  \icarus, 319, 599, \dodoi{10.1016/j.icarus.2018.10.013}

\bibitem[{R.~S. {French} \& M.~R. {Showalter}(2012){French} \&
  {Showalter}}]{fre12}
{French}, R.~S., \& {Showalter}, M.~R. 2012, \bibinfo{title}{{Cupid is doomed:
  An analysis of the stability of the inner uranian satellites},} Icarus, 220,
  911, \dodoi{10.1016/j.icarus.2012.06.031}

\bibitem[{J. {Fuller}(2014){Fuller}}]{ful14}
{Fuller}, J. 2014, \bibinfo{title}{{Saturn ring seismology: Evidence for stable
  stratification in the deep interior of Saturn},} \icarus, 242, 283,
  \dodoi{10.1016/j.icarus.2014.08.006}

\bibitem[{J. {Fuller} {et~al.}(2016){Fuller}, {Luan}, \& {Quataert}}]{ful16}
{Fuller}, J., {Luan}, J., \& {Quataert}, E. 2016, \bibinfo{title}{{Resonance
  locking as the source of rapid tidal migration in the Jupiter and Saturn moon
  systems},} Montly Notices of the Royal Astronomical Society, 458, 3867,
  \dodoi{10.1093/mnras/stw609}

\bibitem[{J. {Fuller} {et~al.}(2026){Fuller}, {Parisi}, {Markham}, {Friedson},
  \& {Fuentes}}]{ful26}
{Fuller}, J., {Parisi}, M., {Markham}, S., {Friedson}, A.~J., \& {Fuentes},
  J.~R. 2026, \bibinfo{title}{{Excitation and Damping of Oscillation Modes in
  Gaseous Planets},} arXiv e-prints, arXiv:2602.12348,
  \dodoi{10.48550/arXiv.2602.12348}

\bibitem[{P. {Goldreich} {et~al.}(1989){Goldreich}, {Murray}, {Longaretti}, \&
  {Banfield}}]{gol89}
{Goldreich}, P., {Murray}, N., {Longaretti}, P.~Y., \& {Banfield}, D. 1989,
  \bibinfo{title}{{Neptune's Story},} Science, 245, 500,
  \dodoi{10.1126/science.245.4917.500}

\bibitem[{M.~M. {Hedman}(2015){Hedman}}]{hed15}
{Hedman}, M.~M. 2015, \bibinfo{title}{{Why Are Dense Planetary Rings Only Found
  between 8 AU and 20 AU?},} ApJL, 801, L33,
  \dodoi{10.1088/2041-8205/801/2/L33}

\bibitem[{M.~M. {Hedman} \& P.~D. {Nicholson}(2013){Hedman} \&
  {Nicholson}}]{hed13}
{Hedman}, M.~M., \& {Nicholson}, P.~D. 2013, \bibinfo{title}{{Kronoseismology:
  Using Density Waves in Saturn's C Ring to Probe the Planet's Interior},} AJ,
  146, 12, \dodoi{10.1088/0004-6256/146/1/12}

\bibitem[{M.~M. {Hedman} \& P.~D. {Nicholson}(2014){Hedman} \&
  {Nicholson}}]{hed14}
{Hedman}, M.~M., \& {Nicholson}, P.~D. 2014, \bibinfo{title}{{More
  Kronoseismology with Saturn's rings},} \mnras, 444, 1369,
  \dodoi{10.1093/mnras/stu1503}

\bibitem[{R.~A. {Jacobson}(2009){Jacobson}}]{jac09}
{Jacobson}, R.~A. 2009, \bibinfo{title}{{The Orbits of the Neptunian Satellites
  and the Orientation of the Pole of Neptune},} AJ, 137, 4322,
  \dodoi{10.1088/0004-6256/137/5/4322}

\bibitem[{E. {Karkoschka}(2003){Karkoschka}}]{kar03}
{Karkoschka}, E. 2003, \bibinfo{title}{{Sizes, shapes, and albedos of the inner
  satellites of Neptune},} Icarus, 162, 400,
  \dodoi{10.1016/S0019-1035(03)00002-2}

\bibitem[{V. {Lainey} {et~al.}(2017){Lainey}, {Jacobson}, {Tajeddine},
  {Cooper}, {Murray}, {Robert}, {Tobie}, {Guillot}, {Mathis}, {Remus},
  {Desmars}, {Arlot}, {De Cuyper}, {Dehant}, {Pascu}, {Thuillot}, {Le
  Poncin-Lafitte}, \& {Zahn}}]{lai17}
{Lainey}, V., {Jacobson}, R.~A., {Tajeddine}, R., {et~al.} 2017,
  \bibinfo{title}{{New constraints on Saturn's interior from Cassini
  astrometric data},} Icarus, 281, 286, \dodoi{10.1016/j.icarus.2016.07.014}

\bibitem[{V. {Lainey} {et~al.}(2020){Lainey}, {Casajus}, {Fuller}, {Zannoni},
  {Tortora}, {Cooper}, {Murray}, {Modenini}, {Park}, {Robert}, \&
  {Zhang}}]{lai20}
{Lainey}, V., {Casajus}, L.~G., {Fuller}, J., {et~al.} 2020,
  \bibinfo{title}{{Resonance locking in giant planets indicated by the rapid
  orbital expansion of Titan},} Nature Astronomy, 4, 1053,
  \dodoi{10.1038/s41550-020-1120-5}

\bibitem[{C.~R. {Mankovich} \& J. {Fuller}(2021){Mankovich} \&
  {Fuller}}]{man21}
{Mankovich}, C.~R., \& {Fuller}, J. 2021, \bibinfo{title}{{A diffuse core in
  Saturn revealed by ring seismology},} Nature Astronomy, 5, 1103,
  \dodoi{10.1038/s41550-021-01448-3}

\bibitem[{M.~S. {Marley} \& C.~C. {Porco}(1993){Marley} \& {Porco}}]{mar93}
{Marley}, M.~S., \& {Porco}, C.~C. 1993, \bibinfo{title}{{Planetary Acoustic
  Mode Seismology: Saturn's Rings},} \icarus, 106, 508,
  \dodoi{10.1006/icar.1993.1189}

\bibitem[{C.~D. Murray \& S.~F. Dermott(1999)Murray \& Dermott}]{md99}
Murray, C.~D., \& Dermott, S.~F. 1999, Solar System Dynamics (Cambridge
  University Press)

\bibitem[{R. {Rufu} \& R.~M. {Canup}(2017){Rufu} \& {Canup}}]{ruf17}
{Rufu}, R., \& {Canup}, R.~M. 2017, \bibinfo{title}{{Triton's Evolution with a
  Primordial Neptunian Satellite System},} AJ, 154, 208,
  \dodoi{10.3847/1538-3881/aa9184}

\bibitem[{M.~R. {Showalter} {et~al.}(2019){Showalter}, {de Pater}, {Lissauer},
  \& {French}}]{sho19}
{Showalter}, M.~R., {de Pater}, I., {Lissauer}, J.~J., \& {French}, R.~S. 2019,
  \bibinfo{title}{{The seventh inner moon of Neptune},} Nature, 566, 350,
  \dodoi{10.1038/s41586-019-0909-9}

\bibitem[{M.~S. {Tiscareno} {et~al.}(2013){Tiscareno}, {Hedman}, {Burns}, \&
  {Castillo-Rogez}}]{tis13}
{Tiscareno}, M.~S., {Hedman}, M.~M., {Burns}, J.~A., \& {Castillo-Rogez}, J.
  2013, \bibinfo{title}{{Compositions and Origins of Outer Planet Systems:
  Insights from the Roche Critical Density},} ApJL, 765, L28,
  \dodoi{10.1088/2041-8205/765/2/L28}

\bibitem[{B. {Wang} {et~al.}(2025){Wang}, {Xi}, {Jianguo}, {Jiawen}, {Xiaowen},
  \& {Wutong}}]{wan25}
{Wang}, B., {Xi}, L., {Jianguo}, Y., {et~al.} 2025, \bibinfo{title}{{A
  Plausible Minimum Value of the Neptunian Tidal Dissipation Factor Estimated
  from Triton's Astrometric Observations},} Solar System Research, 59, 6,
  \dodoi{10.1134/S0038094624601440}

\bibitem[{K. {Zhang} \& D.~P. {Hamilton}(2008){Zhang} \& {Hamilton}}]{zha08}
{Zhang}, K., \& {Hamilton}, D.~P. 2008, \bibinfo{title}{{Orbital resonances in
  the inner neptunian system. II. Resonant history of Proteus, Larissa,
  Galatea, and Despina},} Icarus, 193, 267,
  \dodoi{10.1016/j.icarus.2007.08.024}

\end{thebibliography}
\bibliographystyle{aasjournalv7}



\end{document}